\documentclass[conference]{IEEEtran}
\IEEEoverridecommandlockouts
\usepackage{cite}
\usepackage{amsmath,amssymb,amsfonts}
\usepackage{graphicx}
\usepackage{xcolor}
\def\BibTeX{{\rm B\kern-.05em{\sc i\kern-.025em b}\kern-.08em
		T\kern-.1667em\lower.7ex\hbox{E}\kern-.125emX}}
\usepackage{epsfig}
\usepackage{theorem}
\usepackage{mathrsfs}
\usepackage[utf8]{inputenc}
\usepackage[T1]{fontenc}
\usepackage{newtxtext,newtxmath}
\usepackage{pdfrender}
\usepackage{subfigure}
\usepackage{caption}
\usepackage{csquotes}
\newcommand*{\boldgreek}[1]{%
	\textpdfrender{%
		TextRenderingMode=FillStroke,%
		LineWidth=.35pt,%
	}{#1}%
}
\usepackage{mathrsfs}
\usepackage{euscript}
\usepackage{graphicx}
\usepackage{multirow}
\usepackage{algorithmicx}
\usepackage{algpseudocode}

\usepackage{cite}
\usepackage{url}
\usepackage[export]{adjustbox}
\usepackage{algorithm}
\begin{document}
	
	\title{Joint Transmission in QoE-Driven Backhaul-Aware MC-NOMA Cognitive Radio Network}
	\author{\IEEEauthorblockN {Hosein~Zarini$^{\dag}$, Ata Khalili$^{\S}$, Hina Tabassum$^{\star}$, and Mehdi~Rasti$^{\dag}$	\thanks{This research was supported by a Discovery Grant funded by the Natural Sciences and Engineering Research Council of Canada.}}
		$^{\dag}$Department of Computer Engineering and IT, Amirkabir University of Technology, Tehran, Iran\\
		$^\S$Electronics Research Institute, Sharif University of Technology, Tehran, Iran\\
		$^{\star}$Department of Electrical Engineering and Computer Science at  York  University,  Canada.\\
	}
	\maketitle
	\begin{abstract}
		In this paper, we develop a  resource allocation framework to optimize the downlink transmission of a backhaul-aware multi-cell cognitive radio network (CRN) which is enabled with multi-carrier non-orthogonal multiple access (MC-NOMA). The considered CRN is composed of a single macro base station (MBS) and multiple small BSs (SBSs) that are referred to as the primary and secondary tiers, respectively. For the primary tier, we consider orthogonal frequency division multiple access (OFDMA) scheme and also Quality of Service (QoS) to evaluate the user satisfaction. 
		On the other hand in secondary tier, MC-NOMA is employed and the user satisfaction for web, video and audio as popular multimedia services is
		evaluated by Quality-of-Experience (QoE). Furthermore, each user in secondary tier can be served simultaneously by multiple SBSs over a subcarrier via Joint Transmission (JT). In particular, we formulate a joint optimization problem of power control and scheduling (i.e., user association and subcarrier allocation) in secondary tier to maximize total achievable QoE for the secondary users. An efficient resource allocation mechanism has been developed to handle the non-linear form interference and to overcome the non-convexity of QoE serving functions.
		The scheduling and power control policy leverage on  Augmented Lagrangian Method (ALM).
		Simulation results reveal that proposed solution approach can control the interference and JT-NOMA improves total perceived QoE compared to the existing schemes.
	\end{abstract}
	
	
	\section{Introduction}
	\par Provisioning multimedia services such as video streaming, audio applications, web browsing, file download and best-effort services in forthcoming  wireless networks, prioritizes  user-centric resource allocation over conventional network-centric approach.  In contrast to Quality of Service (QoS) in conventional wireless networks, user-centric resource allocation  considers the users' Quality of Experience (QoE) which is defined as the service-based subjective perception of end users\cite{QoE-2}. 
	To enhance spectral reuse, multi-carrier non-orthogonal multiple access (MC-NOMA) is a potential  technique  that allows multiple users' transmissions over the same subcarrier\cite{NOMA_award, NOMA_MC}. 
	Nevertheless, it is demonstrated in \cite{NOMA_CRN} that MC-NOMA can initiate additional interference and necessitates efficient spectrum allocation. 
	Recently, it is shown that the integration of joint transmission (JT) as a coordinated multi-point (CoMP) feature with MC-NOMA can achieve considerable performance gain\cite{JT_NOMA}. JT stimulates the cooperation of base stations (BSs) to serve a user through multiple BSs over a specified subcarrier.
	However, to realize the JT, backhaul capacity\footnote{Backhaul capacity for a specific BS can be defined as the number of users associated with that BS i.e., the load factor \cite{LC} or as the total transmitting data rate through that BS \cite{backhaul_1}.} of BSs is a limiting factor for serving the users cooperatively \cite{JT2}.  
	\par Recently, QoE has been considered as the user-centric
	criterion  in various research works \cite{QoE_TVT,QoE_TCCN,QoE_TWC}.
	The authors in \cite{QoE_TVT} proposed a resource allocation scheme for web and video  multimedia services using multi-antenna BSs. In \cite{QoE_TCCN}, a resource allocation for satisfying web, video and audio multimedia services is carried out in Device-to-Device (D2D) networks. The authors in \cite{QoE_TWC}  employed MC-NOMA and performed a QoE-driven resource allocation, where user association and subcarrier allocation operations were treated over different phases and each user was  served by only one BS. 
	
	Keeping in view the BS densification in future  wireless networks, we consider JT-NOMA  as a potential technique to enhance the multimedia user perception. To our best knowledge, this integration has not been investigated yet in the context of QoE-aware resource allocation.
	
	\par In this paper, we develop a QoE-aware resource allocation framework to support web surfing, video streaming, and audio application as multimedia services. We propose integrating JT with MC-NOMA in a two-tier multi-cell cognitive radio network (CRN) with a single macro BS (MBS) and multiple small BSs (SBSs) with finite backhauling capacity as the primary and secondary tier respectively. To  enhance spectrum reuse, we consider underlay \text{(UDL)} mode of CRN for sharing a subcarrier between a primary and secondary user and co-channel deployment (CCD), to further share a subcarrier among multiple secondary users. Specifically, to the aim of maximizing total perceived QoE for the secondary users, we perform power control and  joint scheduling (i.e., the joint operation of subcarrier allocation and user association) in secondary tier. We ensure the power constraints for all the BSs, QoS and QoE requirement thresholds for the primary and secondary users respectively and backhaul capacity for the SBSs. 
	Hence, the main contributions of this paper are summarized as follows:
	\begin{itemize}
		\item We investigate JT-NOMA in multi-cell backhaul limited CRN to enhance perceived QoE.
		\item We adopt an efficient joint scheduling in secondary tier among users/SBSs/subcarriers through binary linearizion technique. We propose an efficient power control mechanism by handling the quasi-concave non-linear JT interference based on the Augmented Lagrangian Method (ALM).
		\item Simulation results verify that the proposed resource allocation outperforms the existing literature in terms of perceived QoE. 
	\end{itemize}

	\section{System Model and Assumptions}
\begin{figure}[t]
		\hspace{-0.3cm}\includegraphics[width=9.00cm,height=5.500cm]{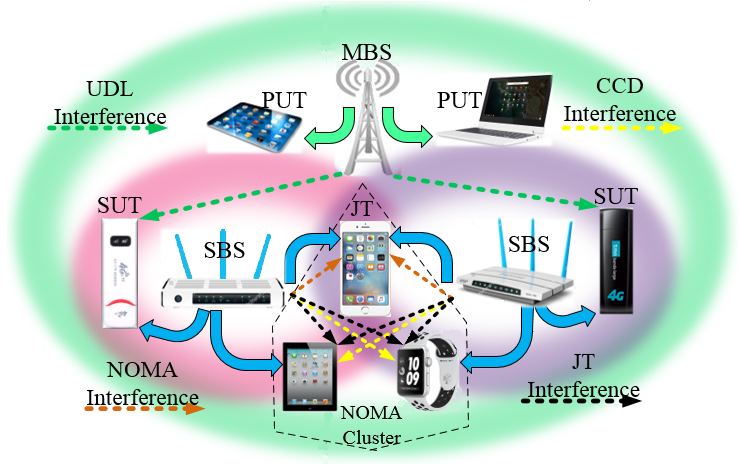}
		\caption{\small JT-NOMA CRN multimedia network.}
		\label{Fig_1}
	\end{figure}

	As shown in Fig. \ref{Fig_1}, we consider downlink CRN where users and multiple SBSs are equipped with single antenna and uniformly distributed within the coverage area of single MBS.
	The primary tier consists of a cell centered MBS with wide coverage area and unlimited backhaul capacity denoted by $s$ and set of primary user terminals (PUTs) denoted by $\mathcal{M}=\{1,...,M\}$. The assessment metric for the PUTs is assumed to be conventional QoS and the subcarrier allocation in primary tier is predefined based on OFDMA. The secondary tier consists set of SBSs denoted by $\mathcal{L}=\{1,...,L\}$ underlaying the MBS and the set of secondary user terminals (SUTs) given by $\mathcal{G}=\{1,...,G\}$ with MC-NOMA for the subcarrier allocation and QoE as the assessment metric. The backhaul limited SBSs with limited coverage area are connected with each other and with MBS through digital subscriber line or cable modem. The secondary tier employs universal spectrum reuse where every SBS shares all  subcarriers denoted by $\mathcal{N}=\{1,2,..,N\}$. 
	All the BSs work on licensed spectrum and the secondary tier thus, experiences following interference:
	\begin{itemize}
		\item {\bf UDL Interference:} results from sharing a subcarrier among a primary and a secondary user.  The MBS and SBSs impose inter-tier UDL interference to the involving SUTs and PUTs, respectively.
		\item  {\bf CCD Interference:} results from the reuse of a subcarrier among multiple secondary users associated with different SBSs. In particular, SUTs experience intra-tier inter-cell interference corresponding to adjacent interfering SBSs.
		\item {\bf NOMA Interference:} Since MC-NOMA is exploited in secondary tier,  a subcarrier may be shared among multiple SUTs within a SBS yielding intra-tier intra-cell interference.
		\item {\bf JT Interference:} The secondary tier is also JT-enabled, resulting the joint interference from SBSs involving in JT operation to other SUTs sharing the same JT subcarrier.
	\end{itemize}
	Finally, each SUT can be served simultaneously over multiple subcarriers within a cell using carrier aggregation. 
	Let us denote $h^n_{x,y}$ as the channel gain between BS $x$ and user $y$ over the subcarrier $n$ with quasi-static, independent and identically distributed (i.i.d) Rayleigh model. Also we denote $p^{n}_{l,g}$ as the downlink transmit power of SBS $l$ to SUT $g$ and $q^{n}_{m}$ as the downlink transmit power of the MBS to PUT $m$ both over the subcarrier $n$ with corresponding matrices ${\textrm{p}} = \Big[{p^{n}_{l,g}}\Big]_{{|\mathcal{L}|}\times{|\mathcal{G}|}\times{|\mathcal{N}|}}$ and ${\textrm{q}} = \Big[{q^{n}_{m}}\Big]_{{|\mathcal{M}|}\times{|\mathcal{N}|}}$, respectively. 
	Respectless to scheduling operation, the received signal of SUT $g$ from SBS $l$ over the subcarrier $n$ is given by (\ref{SUT_signal}), where $x_{a,b}^{c}$ indicates the transmitted signal from BS $a$ to SUT $b$ over the subcarrier $c$ with unit power (i.e., $\mathbb{E}\{{|{x}_{a,b}^{c}}|^2\}$ = 1). We denote $\omega_{g}$ as the zero mean Additive White Gaussian Noise (AWGN) for SUT $g$ with the variance of $\sigma_{g}^{2}$ and $\phi_{l^{\prime}}$ and $\phi_{l^{\prime\prime}}$ are the phases of the two received interfering signals\cite{JT7}.
		\begin{align}\label{SUT_signal}
		y_{l,g}^{n} = &\underbrace{\sqrt{p_{l,g}^{n}}h_{l,g}^{n}x_{l,g}^{n}}_{\textrm{Intended Signal}}+\underbrace{\sum_{m\in{\mathcal{M}}}\sqrt{q_{s,m}^{n}}h_{s,g}^{n}x_{s,m}^{n}}_{{\textrm{UDL Interference}}}+\underbrace{\omega_{g}}_{\textrm{AWGN}}\\\nonumber+&\underbrace{\sum_{\substack{l^\prime\ne{l}\\l^\prime\in{\mathcal{L}}}}\,\sum_{\substack{g^\prime\ne{g}\\g^\prime\in{\mathcal{G}}}}\sqrt{p_{l^\prime,g^\prime}^{n}}h_{l^\prime,g}^{n}x_{l^\prime,g^\prime}^{n}}_{{\textrm{\text{CCD} Interference}}}+\underbrace{\sum_{\substack{g^\prime\ne{g}\\g^\prime\in{\mathcal{G}}}}\sqrt{p_{l,g^\prime}^{n}}h_{l,g}^{n}x_{l,g^\prime}^{n}}_{{\textrm{NOMA Interference}}}\\\nonumber+& \underbrace{\sum_{l^\prime\in{\mathcal{L}}}\,\sum_{\substack{l^{\prime\prime}\ne{l^\prime}\\l^{\prime\prime}\in{\mathcal{L}}}}\,\sum_{\substack{g^\prime\ne{g}\\g^\prime\in{\mathcal{G}}}}2\sqrt{p_{l^\prime,g^\prime}^{n}}h_{l^\prime,g}^{n}
			x_{l^\prime,g^\prime}^{n}\sqrt{p_{l^{\prime\prime},g^\prime}^{n}}h_{l^{\prime\prime},g}^{n}x_{l^{\prime\prime},g^\prime}^{n}\mathrm{cos}(\phi_{l^\prime}-\phi_{l^{\prime\prime}})}_{{\textrm{JT Interference}}}.\nonumber
		\end{align}
	\par We assume that the full channel state information (CSI) is available  for the BSs \cite{khalili2}, and channel gains satisfy the order of $|h_{l,1}^{n}|\le{| h_{l,2}^{n}|}\le{...}\le{| h_{l,G}^{n}|}$. Let us denote the set of SUTs in NOMA cluster of subcarrier $n$ at the SBS $l$ by $\mathcal{G}^{n}_{l}$. For specific couple of SUTs in same NOMA cluster e.g., $(g,g^\prime)\in{\mathcal{G}^{n}_{l}}$, the transmitted signal for $g^\prime$ is detectable in $g$ only if $g^\prime$ is the precedent one in the order. Then, $g^\prime$ decodes its own signal while treating $g$-th signal as interference. On the other side, $g$ with better channel condition, decodes $g^\prime$-th superimposed signal first and removes it through Successive Interference Cancellation (SIC) before decoding its own message \cite{NOMA_award}. The procedure is supposed to be performed for every pair of SUTs in same NOMA cluster.
	
	\section{QoE Model and Problem Formulation}
	Based on what discussed so far, the downlink received signal-to-interference-plus-noise-ratio (SINR) for SUT $g$ associated with SBS $l$ over subcarrier $n$ is expressed as
	\begin{align}\label{SINR}
	&\gamma^{n}_{l,g}(\theta,\varepsilon,{\textrm{p}},{\textrm{q}})= \nonumber\\&\frac{p_{l,g}^{n}{|h^{n}_{l,g}}|^{2}}{{{\mathcal{I}_{l,g}^{n}}\text{(UDL)}}+{{\mathcal{I}_{l,g}^{n}}(\text{CCD})}+
		{{\mathcal{I}_{l,g}^{n}}(\text{NOMA)}}+{{\mathcal{I}_{l,g}^{n}}(JT)}+\sigma_{g}^{2}},
	\end{align}
	where,
	\begin{align}
	&{{\mathcal{I}_{l,g}^{n}}\text{(UDL)}}=\sum_{m\in{\mathcal{M}}}\pi_{m}^{n}q_{s,m}^{n}{|h_{s,g}^{n}|}^{2},\\
	&{{\mathcal{I}_{l,g}^{n}}(\text{CCD})}=\sum_{\substack{l^\prime\ne{l}\\l^\prime\in{\mathcal{L}}}}\,\sum_{\substack{g^\prime\ne{g}\\g^\prime\in{\mathcal{G}}}}\theta_{l^\prime,g^\prime}\varepsilon_{l^\prime,g^\prime}^{n} p_{l^\prime,g^\prime}^{n}{|h_{l^\prime,g}^{n}|}^{2},\\
	&{{\mathcal{I}_{l,g}^{n}}(\text{NOMA})}=\sum_{\substack{g^\prime\in{\mathcal{G}^{n}_{l}}\\g^\prime=g+1}}\theta_{l,g^\prime}\varepsilon_{l,g^\prime}^{n}p_{l,g^\prime}^{n}{|h_{l,g}^{n}|}^{2},\\
	&{{\mathcal{I}_{l,g}^{n}}(JT)}=\\\nonumber&\sum_{l^\prime\in{\mathcal{L}}}\,\sum_{\substack{l^{\prime\prime}\ne{l^\prime}\\l^{\prime\prime}\in{\mathcal{L}}}}\,\sum_{\substack{g^\prime\ne{g}\\g^\prime\in{\mathcal{G}}}}2\theta_{l^\prime,g^\prime}\varepsilon_{l^\prime,g^\prime}^{n}p_{l^\prime,g^\prime}^{n}\theta_{l^{\prime\prime},g^\prime}\varepsilon_{l^{\prime\prime},g^\prime}^{n}p_{l^{\prime\prime},g^\prime}^{n}{|h_{l^\prime,g}^{n}|}^{2}{{|h_{l^{\prime\prime},g}^{n}|}}^{2},
	\end{align}
	and $\pi_{m}^{n}$ is the predefined primary tier binary subcarrier allocation variable with the value of 1 if the subcarrier $n$ is allocated to PUT $m$ and zero otherwise. Besides, the binary variables ${\theta} = \Big[{\theta_{l,g}}\Big]_{{|\mathcal{L}|}\times{|\mathcal{G}|}}$ and ${\varepsilon} = \Big[{\varepsilon^{n}_{l,g}}\Big]_{{|\mathcal{L}|}\times{|\mathcal{G}|}\times{|\mathcal{N}|}}$ indicate associating of user $g$ with BS $l$ and allocating the subcarrier $n$ of BS $l$ to user $g$, respectively. Moreover, ${{\mathcal{I}_{l,g}^{n}}(\text{NOMA})}$
	reveals the precedence of SUTs with poor channel condition in decoding order through applying successful SIC. The procedure also stands for the special case where
	couple of JT-enabled and single-associated SUTs (i.e.,
	the SUTs served only by one SBS) are in a NOMA cluster. In ordering of SUTs in a NOMA cluster for demonstrated two-tier CRN in Fig. (\ref{Fig_1}), the cell-edge JT-NOMA SUTs are precedent to cell-centered non-JT-NOMA SUTs\footnote{For the special case where multiple JT-enabled SUTs coexist within a NOMA cluster\cite{JT_NOMA}, a fixed decoding order of users is considered based on average distance of users with all the BSs involving in JT.}. In a NOMA cluster, neighbouring SUTs experience intra-tier intra-cell interference as the result of superposition coding in power domain of the transmitter.
	The achievable data rate of SUT $g$ then is given by
	\begin{equation} \label{Rate_SUE}
	\begin{aligned}
	&{\textrm{R}_{g}(\theta,\varepsilon,{\textrm{p}},{\textrm{q}})}=\sum_{l\in{\mathcal{L}}}\,\sum_{n\in{\mathcal{N}}}\theta_{l,g}\varepsilon_{l,g}^{n}\log_2\big(1+\gamma^{n}_{l,g}(\theta,\varepsilon,{\textrm{p}},{\textrm{q}})\big).\end{aligned}
	\end{equation}
	In this paper, we adopt Mean Opinion Score (MOS) as the main criterion of QoE for web surfing, video streaming, and audio application services with indices $j \in \{1,2,3\}$,~respectively \cite{QoE_TCCN}. 
	\par Let us define the total perceived QoE for the SUTs \cite{QoE_TWC}
	\begin{equation} \label{U_QoE}
	\begin{aligned}
	&U_{QoE}^{j}(\boldgreek{\theta},\boldgreek{\varepsilon},\textbf{\textrm{p}},\textbf{\textrm{q}})=\sum_{g\in{\mathcal{G}}}\textrm{MOS}^{j}_{g}(\boldgreek{\theta},\boldgreek{\varepsilon},\textbf{\textrm{p}},\textbf{\textrm{q}})=\\&\sum_{g\in{\mathcal{G}}}\textrm{MOS}^{j}_{g}\Big(\sum_{l\in{\mathcal{L}}}\sum_{n\in{\mathcal{N}}}\theta_{l,g}\varepsilon_{l,g}^{n}\log_{2}\big(1+\gamma_{l,g}^{n}(\boldgreek{\theta},\boldgreek{\varepsilon},\textbf{\textrm{p}},\textbf{\textrm{q}})\big)\Big),
	\end{aligned}
	\end{equation}
	as the objective function of optimization for the $j$-th QoE service with ($\boldgreek{\theta},\boldgreek{\varepsilon}$) and $(\textbf{\textrm{p}},\textbf{\textrm{q}})$ as the decision variables.
	For the transmit power thresholds we have
	\begin{equation} \label{C1}
	\begin{aligned}
	\sum_{m\in{\mathcal{M}}}\,\sum_{n\in{\mathcal{N}}}\pi_{m}^{n}q_{m}^{n}\le {{q}^{max}},
	\end{aligned}
	\end{equation}
	\begin{equation} \label{C2}
	\begin{aligned}
	\sum_{g\in{\mathcal{G}}}\,\sum_{n\in{\mathcal{N}}}\varepsilon_{l,g}^{n}p_{l,g}^{n}\le {{p}^{max}_{l}} ~~\forall {l},
	\end{aligned}
	\end{equation}
	\\where ${q}^{max}$ and ${p}^{max}_{l}$ are the maximum transmit power for MBS and SBSs, respectively.
	The QoS requirement of PUTs can be ensured if
	\begin{equation} \label{C3}
	\begin{aligned}
	&\tilde{R}_{m}(\boldgreek{\theta},\boldgreek{\varepsilon},\textbf{\textrm{p}},\textbf{\textrm{q}})=\\&\sum_{n\in{\mathcal{N}}}\pi_{m}^{n}\log_2\big(1+\frac{q^{n}_{m}{|h^{n}_{s,m}|}^{2}}{ \sum_{l\in{\mathcal{L}}}\sum_{g\in{\mathcal{G}}} \theta_{l,g}\varepsilon_{l,g}^{n}p_{l,g}^{n}|{h^{n}_{l,m}}|^{2}+\sigma_{g}^2}\big)\ge{\tilde{R}^{min}_{m}}.
	\end{aligned}
	\end{equation} 
	After achieving successful SIC and signal decoding operation in SUTs, the QoE requirement is conceivable if\cite{QoE_TCCN}
	\begin{equation} \label{C4}
	\begin{aligned}
	&\textrm{MOS}^{j}_{g}\Big(\sum_{l\in{\mathcal{L}}}\sum_{n\in{\mathcal{N}}}\theta_{l,g}\varepsilon_{l,g}^{n}\log_{2}\big(1+\gamma_{l,g}^{n}(\boldgreek{\theta},\boldgreek{\varepsilon},\textbf{\textrm{p}},\textbf{\textrm{q}})\big)\Big)\ge{\textrm{MOS}_{g}^{min,j}}.
	\end{aligned}
	\end{equation}
	The backhaul capacity limit for the SBS $l$ can be ensured through 
	\begin{equation} \label{C5}
	\begin{aligned} &C_{l}(\boldgreek{\theta},\boldgreek{\varepsilon},\textbf{\textrm{p}},\textbf{\textrm{q}})=\sum_{g\in{\mathcal{G}}}\sum_{n\in{\mathcal{N}}}\theta_{l,g}\varepsilon_{l,g}^{n}\log_{2}(1+\gamma_{l,g}^{n}(\boldgreek{\theta},\boldgreek{\varepsilon},\textbf{\textrm{p}},\textbf{\textrm{q}}))\le{C^{max}_{l}}.~~
	\end{aligned}
	\end{equation}
	For the backhaul limit as the load of each SBS in terms of number of associated SUTs we have \cite{LC}
	\begin{align} \label{C6}
	\sum_{g\in{\mathcal{G}}}\,\theta_{l,g}\le {Z}_{l} ~~\forall {l}.
	\end{align}
	Regarding  the system feasibility, each SUT is imposed to be associated with one SBS and allocated one subcarrier at least given by
	\begin{align} \label{C7}
	&\sum_{l\in{\mathcal{L}}}\,\theta_{l,g}\ge {1} ~~\forall {g},
	\end{align}
	\begin{align} \label{C8}
	&\sum_{l\in{\mathcal{L}}}\sum_{n\in{\mathcal{N}}}\,\varepsilon^{n}_{l,g}\ge {1} ~~\forall {g},
	\end{align}
	respectively. The complexity of SIC decoding over a specified subcarrier on receiver can be controlled through limiting the number of multiplexed users over that subcarrier as
	\begin{align} \label{C9}
	\sum_{l\in{\mathcal{L}}}\sum_{g\in{\mathcal{G}}}\,\varepsilon_{l,g}^{n}\le {\Omega_{n}} ~~\forall {n},
	\end{align}
	with $\Omega_{n}$ denoting the maximum number of multiplexed users over subcarrier $n$.
	The precedence of the cell association procedure to the subcarrier allocation \cite{CA2} can indicated by 
	\begin{equation} \label{C10}
	\begin{aligned}
	\varepsilon_{l,g}^{n}\le {\theta_{l,g}} ~~\forall {l,g,n}.
	\end{aligned}
	\end{equation}
	The problem at hand for joint QoE-aware optimization of
	the $j$-th service is formally stated as
	\begin{equation}
	\label{QoE_max_prob.1}
	\begin{aligned}
	\mathcal{P}_{1}:\max_{\boldgreek{\theta},\boldgreek{\varepsilon},\textbf{\textrm{p}},\textbf{\textrm{q}}}&~U_{QoE}^{j}(\boldgreek{\theta},\boldgreek{\varepsilon},\textbf{\textrm{p}},\textbf{\textrm{q}})
	\\&\text{s.t.:}~(\ref{C1})-(\ref{C10}),
	\end{aligned}
	\end{equation}
	which is complex due to the objective function along with non-linear and non-convex constraints (\ref{C2})-(\ref{C5}).
	\section{Power Control and Joint Scheduling Algorithm}
	In this section, we provide an alternative decomposition method to address (\ref{QoE_max_prob.1}). In doing so, two sub-problems namely power control and joint scheduling are solved in a joint convergent manner based on Algorithm 1.
	\subsection{Power Control Policy}
	Under predetermined values of variables $[\theta^{(t-1)},\varepsilon^{(t-1)}]$, the joint primitive optimization problem (\ref{QoE_max_prob.1}) is transformed into power control sub-problem with decision variables $\textbf{\textrm{p}}^{(t)}$ and $\textbf{\textrm{q}}^{(t)}$ given by
	\begin{equation}
	\label{QoE-PC_max_min_prob}
	\begin{aligned}
	\mathcal{P}_{2}:\max_{{\textbf{\textrm{p}}}^{(t)},\textbf{\textrm{q}}^{(t)}}&~U_{QoE}^{j}({\theta}^{(t-1)},{\varepsilon}^{(t-1)},\textbf{\textrm{p}}^{(t)},\textbf{\textrm{q}}^{(t)})
	\\&\text{s.t.:}~(\ref{C1})-(\ref{C5}).
	\end{aligned}
	\end{equation}
	The reformulated sub-problem is non-linear and quasi-concave form because of JT interference (\ref{SINR}). For multiplication of two continuous variables ${p_{l^\prime,g^\prime}^{n}}^{(t)}$ and ${p_{l^{\prime\prime},g^\prime}^{n}}^{(t)}$, the following convex inequality for any fixed $\lambda
	\textgreater{0}$ holds\footnote{Recall that for $\lambda=\dfrac{{p_{l^{\prime\prime},g^\prime}^{n}}^{(t)}}{{{p_{l^\prime,g^\prime}^{n}}^{(t)}}}$, the approximation is tight.}\cite{QoE_TVT}
	\begin{equation} \label{lambda}
	\begin{aligned}
	&{{p_{l^\prime,g^\prime}^{n}}^{(t)}{p_{l^{\prime\prime},g^\prime}^{n}}}^{(t)}\le{\dfrac{\lambda}{2}{({{p_{l^\prime,g^\prime}^{n}}}^{(t)})}^{2}+\dfrac{1}{2\lambda}{{(p_{l^{\prime\prime},g^\prime}^{n}}^{(t)})}^{2}}~~~\\&\forall{{l^{\prime}},{l^{\prime\prime}},g\prime,n},{l^{\prime}}\ne{{l^{\prime\prime}}}.
	\end{aligned}
	\end{equation}
	Then, the convex form of JT interference can be given as
	\begin{eqnarray}
	\begin{aligned}
	&{\mathcal{I}_{l,g}^{n}}({JT})=\\&\sum_{l^\prime\in{L}}\,\sum_{\substack{l^{\prime\prime}\ne{l^\prime}\\l^{\prime\prime}\in{L}}}\,\sum_{\substack{g^\prime\ne{g}\\g^\prime\in{G}}}2[\dfrac{\lambda}{2}{\theta_{l^\prime,g^\prime}}^{{(t-1)}}{\varepsilon_{l^\prime,g^\prime}^{n}}^{{(t-1)}}{({{p_{l^\prime,g^\prime}^{n}}}^{(t)})}^{2}{|h_{l^\prime,g}^{n}|}^{2}+\\&\dfrac{1}{2\lambda}{\theta_{l^{\prime\prime},g^\prime}}^{{(t-1)}}{\varepsilon_{l^{\prime\prime},g^\prime}^{n}}^{{(t-1)}}{{(p_{l^{\prime\prime},g^\prime}^{n}}^{(t)})}^{2}{|h_{l^{\prime\prime},g}^{n}|}^{2}].
	\end{aligned}
	\end{eqnarray}
	Due to incorporating interference terms in (\ref{C3})-(\ref{C5}), the sub-problem (\ref{QoE-PC_max_min_prob}) is still non-convex which initiates a duality gap between the solution of (\ref{QoE-PC_max_min_prob}) and its dual problem. We employ a power control policy leveraging ALM \cite{ALM1} where $\alpha$ is the augmenting Lagrangian multiplier and 
	$\Psi_b-\Psi_f$ are the Lagrangian multipliers for the corresponding constraints.  
	Accordingly, the dual problem of (\ref{QoE-PC_max_min_prob}) for the $j$-th service is expressed as following
	\begin{equation}\label{dual_1}
	D_{\mathcal{P}_{2}}^{j} = \min_{\alpha,\Psi_b,...,\Psi_f}\max_{\textbf{\textrm{p}},\textbf{\textrm{q}}}\zeta_{\mathcal{P}_{2}}^{j}(\alpha,\Psi_b,...,\Psi_f,\textbf{\textrm{p}},\textbf{\textrm{q}}).
	\end{equation}
	To address (\ref{dual_1}) through ALM, the Lagrangian function can be given by (\ref{Lagrange}) on top of the next page.
	\begin{table*}
		\begin{equation}
		\label{Lagrange}
		\begin{aligned}
		&\zeta_{\mathcal{P}_{2}}^{j}(\Psi_b,...,\Psi_f,{\textbf{\textrm{p}}},\textbf{\textrm{q}})=U_{QoE}^{j}({\theta}^{(t-1)},{\varepsilon}^{(t-1)},\textbf{\textrm{p}}^{(t)},\textbf{\textrm{q}}^{(t)})+\frac{1}{2\alpha^{(t)}}\Bigg[\Bigg(\Bigg(\Bigg[\Psi_b+\alpha^{(t)} \Big(\sum_{m\in\mathcal{M}}\,\sum_{n\in\mathcal{N}}\,{q_{m}^{n}}^{(t)}- {{q}^{max}}\Big)\Bigg]^{+}\Bigg)^{2}-{\Psi_b}^{2}\Bigg)+ \\&\sum_{l\in{\mathcal{L}}}\Bigg(\Bigg(\Bigg[\Psi_{c_{l}}+\alpha^{(t)} \Big(\sum_{g\in\mathcal{G}}\,\sum_{n\in\mathcal{N}}\,{\varepsilon_{l,g}^{n}}^{(t-1)}{p_{l,g}^{n}}^{(t)}- {{p}_{l}^{max}}\Big)\Bigg]^{+}\Bigg)^{2}-{\Psi_{c_{l}}}^{2}\Bigg)+
		\sum_{l\in{\mathcal{L}}}\Bigg(\Bigg(\Bigg[\Psi_{d_{l}}+\alpha^{(t)} \Big(\,C_{l}(\theta^{(t-1)},\varepsilon^{(t-1)},\textbf{\textrm{p}}^{(t)},\textbf{\textrm{q}}^{(t)})-{C}_{l}^{max}\Big)\Bigg]^{+}\Bigg)^{2}-\Psi_{d_{l}}^{2}\Bigg)+ \\&\sum_{m\in{\mathcal{M}}}\Bigg(\Bigg(\Bigg[\Psi_{e_{m}}+\alpha^{(t)} \Big(\,{\tilde{R}^{min}_{m}}-\tilde{R}_{m}(\theta^{(t-1)},\varepsilon^{(t-1)},\textbf{\textrm{p}}^{(t)},\textbf{\textrm{q}}^{(t)})\Big)\Bigg]^{+}\Bigg)^{2}-\Psi_{e_{m}}^{2}\Bigg)+ 
		\sum_{g\in{\mathcal{G}}}\Bigg(\Bigg(\Bigg[\Psi_{f_{g}}+\alpha^{(t)} \Big(\,{\textrm{\scriptsize{MOS}}^{min,j}_{g}}-\textrm{\scriptsize{MOS}}_{g}^{j}(\theta^{(t-1)},\varepsilon^{(t-1)},\textbf{\textrm{p}}^{(t)},\textbf{\textrm{q}}^{(t)}) \Big)\Bigg]^{+}\Bigg)^{2}-\Psi_{f_{g}}^{2}\Bigg)\Bigg].
		\end{aligned}
		\end{equation}
		\hrule
	\end{table*}
	By updating the Lagrangian multipliers following \cite{khalili2} after some iterations, the duality gap will be zero, Karush–Kuhn–Tucker (KKT) conditions hold which yields a sub-optimal solution\cite{Schuber}.
	\subsection{Joint scheduling}
	Given $[{\textrm{p}}^{(t)},{\textrm{q}}^{(t)}]$ matrices obtained from power control policy, constraints (\ref{C3})-(\ref{C5}) in (\ref{QoE_max_prob.1}) along with the objective function are in non-linear multiplicative form. We define new binary decision variable as ${\chi_{l,g}^{n}}^{(t)} ={\theta_{l,g}}^{(t)}{\varepsilon_{l,g}^{n}}^{(t)}$ indicating the joint operation of scheduling. By including the following additional constraints \cite{optimization}, 
	\begin{align}\label{C11}
	&{\chi_{l,g}^{n}}^{(t)}\le {\theta_{l,g}}^{(t)} ~~\forall {l,g,n},\\
	\label{C12}&{\chi_{l,g}^{n}}^{(t)}\le {\varepsilon_{l,g}^{n}}^{(t)} ~~\forall {l,g,n},\\
	\label{C13}
	&{\theta_{l,g}}^{(t)}+{\varepsilon_{l,g}^{n}}^{(t)}-1\le {\chi_{l,g}^{n}}^{(t)} ~~\forall {l,g,n},
	\end{align}
	the non-linear form arised from multiplication of binary variables is tackled hitherto \cite{WKh}. The non-convexity due to interference terms incorporation nonetheless, is the challenging issue yet. 
	Hence, the relaxation of binary decision variables into continuous domain is invoked so as to make a more tractable sub-problem. Further, binary-forcing constraints \cite{Khalili,TWC_Ata,Khalili_G1,Khalili_G2} given by 
	\begin{align}
	&\sum_{l\in{\mathcal{L}}}\sum_{g\in{\mathcal{G}}}\sum_{n\in{\mathcal{N}}}\Big[\big({\varepsilon_{l,g}^{n}}^{(t)}\big)-\big({{\varepsilon_{l,g}^{n}}^{(t)}}\big)^{2}\Big]\le{0},\\
	&\sum_{l\in{\mathcal{L}}}\sum_{g\in{\mathcal{G}}}\Big[\big({\theta_{l,g}}^{(t)}\big)-\big({{\theta_{l,g}}^{(t)}}\big)^{2}\Big]\le{0},\\
	&\sum_{l\in{\mathcal{L}}}\sum_{g\in{\mathcal{G}}}\sum_{n\in{\mathcal{N}}}\Big[\big({\chi_{l,g}^{n}}^{(t)}\big)-\big({{\chi_{l,g}^{n}}^{(t)}}\big)^{2}\Big]\le{0},
	\end{align}
	can guarantee the binary domain of decision variables.
	The feasible region for the aforementioned constraints consists of corner critical points i.e., \{0,1\}. 
	Thus, the joint optimization problem for the $j$-th service (\ref{QoE_max_prob.1}) is restated as
	\begin{equation}
	\label{QoE_max_prob.2}
	\begin{aligned}
	\mathcal{P}_{3}:\max_{\boldgreek{\chi}^{(t)}\boldgreek{\theta}^{(t)},\boldgreek{\varepsilon}^{(t)}}& U_{QoE}^{j}(\boldgreek{\theta}^{(t)},\boldgreek{\varepsilon}^{(t)},\textrm{p}^{(t)},\textrm{q}^{(t)})
	\\&\text{s.t.: (\ref{C2})-(\ref{C10}), (\ref{C11})-(\ref{C13})}.
	\end{aligned}
	\end{equation}
	Reformulated optimization problem (\ref{QoE_max_prob.2}) is still non-convex due to incorporating decision variable $\chi$ as the indicator of interference. 
	The remaining solution for the sub-problem is completely similar to power control policy henceforth where, ALM with respect to decision variables $\chi^{(t)},\theta^{(t)},\varepsilon^{(t)}$ is applied to obtain the sub-optimal solution.
	Then, dual problem will be indicated by
	\begin{eqnarray}
	\label{dual_2}
	\begin{aligned}
	D_{\mathcal{P}_{3}}^{j}=
	\min_{\alpha,\Delta_b,...,\Delta_v}\max_{\boldgreek{\chi},\boldgreek{\theta},\boldgreek{\varepsilon}}\zeta_{\mathcal{P}_{3}}^{j}(\alpha,\Delta_b,...,\Delta_v,\boldgreek{\chi},\boldgreek{\theta},\boldgreek{\varepsilon}),
	\end{aligned}
	\end{eqnarray}
	where, $\Delta_b,...,\Delta_v$ are the corresponding Lagrangian multipliers. 
	The Lagrangian multipliers can be updated as in \cite{khalili2} to obtain the sub-optimal solution.
	\begingroup
	\begin{algorithm}[t]\label{JRA}
		\footnotesize{
			\caption{Joint Resource Allocation}
			\textbf{Initialization}: $Err=10^{-3}$, ${\chi}^{(0)},{\theta}^{(0)}$ and ${\varepsilon}^{(0)}$ are randomly initialized, transmit power matrices are initialized across subcarriers (i.e., ${p_{l,g}^{n}}^{(0)}=p^{max}_{l}/N~\forall{m,n}, {q_{m}^{n}}^{(0)}=q^{max}/N~\forall{m,n}$. Iteration numbers t and T are initialized to 1, ALM Convergence = Final Convergence = \textbf{false}, $\alpha$ = 2 and other Lagrangian multipliers are set to 0.1.
			\begin{algorithmic}[1]
				\Repeat
				\Repeat
				\State Solve (\ref{QoE-PC_max_min_prob}) to get ${\textrm{p}}^{(t)}$ and ${\textrm{q}}^{(t)}$.
				\If{$\begin{cases}
					{|{p}}^{(t)}-{\textrm{p}}^{(t-1)}| < Err
					\\ {|\textrm{q}}^{(t)}-{\textrm{q}}^{(t-1)}| < Err
					\end{cases}$}
				\State ALM Convergence = \textbf{true}.  
				\Else
				\State $t=t+1$.
				\State $\textrm{Update Lagrangian multipliers }$ $(\Psi_b)-(\Psi_f)$.
				\EndIf
				\Until{ALM Convergence}
				\State $t$=1.
				\Repeat
				\State Solve (\ref{QoE_max_prob.2}) to get $\theta^{(t)},\varepsilon^{(t)}$ and $\chi^{(t)}$.
				\If
				{${\chi}^{(t)}={\chi}^{(t-1)}$}
				\State ALM Convergence = \textbf{true}.    
				\Else
				\State $\textrm{Update Lagrangian multipliers }$ $(\Delta_b)-(\Delta_v)$.
				\State $t=t+1$.
				\EndIf
				\Until{ALM Convergence}
				\State Calculate ${U}^{j^{(T)}}_{QoE}\big(\theta,\varepsilon,{\textrm{p}},{\textrm{q}}\big)$.
				\If{${|{U}^{j^{(T)}}_{QoE}}(\theta,\varepsilon,{\textrm{p}},{\textrm{q}})-{U^{j^{(T-1)}}_{QoE}}(\theta,\varepsilon,{\textrm{p}},{\textrm{q}})|< Err$
				}
				\State Final Convergence = \textbf{true}. 
				\Else
				\State $T=T+1$.
				\EndIf
				\Until{Final Convergence}
		\end{algorithmic}}
	\end{algorithm}
	\endgroup
	\section{Simulation Results}
	In this section, simulation results are presented to investigate the efficiency of proposed solutions of the joint optimization problem. The MBS covers 500 meters of area, coexisting with SBSs with 50 meters of coverage. Maximum transmit power of the MBS ($q^{max}$) and all the SBSs ($p^{max}_{l}~\forall{l}$) are 42dBm and 37dBm, respectively. The bandwidth of 15 kHz for each subcarrier is considered. For the backhaul capacity, the commercial optical fiber modem  capable of supporting $C^{max}_{l} = 11.183$~Mbps $\forall{l}$ is considered \cite{Schuber}. Other simulation parameters are set as $\sigma^2$ = -117 dBm, $\tilde{R}^{min}_{m} = 2$~bps/Hz $\forall{m}$, $\textrm{MOS}^{j,min}_{g} = 1$~bps/Hz $\forall{g,j\in\{1,2,3}\}$, $Z_{l} =
	3~\forall{l}$ and $\Omega_{n} =
	2~\forall{n}$, unless specified. 
	
	The convergence behavior of the proposed solution is demonstrated with average MOS and average data rate on left and right hand side of the figures, respectively.~Also, the performance evaluation is demonstrated in figures through four cases, i.e., proposed JT-NOMA, non-JT-NOMA according to \cite{QoE_TWC}, as well as   Orthogonal Multiple Access (OMA) schemes like JT-OMA and non-JT-OMA for benchmarking purposes.
	For all the services, minimum target MOS i.e., $\textrm{MOS}_{g}^{min,1}$ is mapped to $R^{min}_{g}=2$~bit/s/Hz while the maximum MOS value, 5 for web service and 4.5 for video and audio services are mapped to $R^{max}_{g}=7$~bit/s/Hz\cite{QoE_TVT}.
	\subsubsection{Web Service Assessment}
	\par The network dependent parameters consisting of maximum segment size, round trip time and web page size are set as in \cite{QoE_TWC}. The network configuration parameters are  $L$ = 10, $N$ = 32, and $M$ = 6. Fig. \ref{fig:Comp}(a) is related to the convergence behavior of our proposed solution for different numbers of SUTs indicating the exact values of load latency \cite{QoE_TWC} at convergence point. In Fig. \ref{fig:Comp}(d) with increasing number of SUTs, 'JT-NOMA' scheme outperforms others in terms of average MOS. Also, \text{CCD}, NOMA and JT interference terms, result in decreasing trend of average MOS per SUT versus increasing the number of SUTs. 
	\subsubsection{Video Service Assessment}
	For the video service with common video coding H.264/AVC, the network configuration parameters are $L$ = 10, $N$ = 16, and $G$ = 10.
	In single-snapshot scenario shown in Fig. \ref{fig:Comp}(b), convergence of the proposed algorithm and corresponding peak signal to noise ratio (PSNR) values \cite{QoE_TVT} at convergence point are demonstrated for different values for PUTs. The multi-snapshot Fig. 2(e) demonstrates decreasing average MOS per SUT versus increasing the number of PUTs due to higher UDL interference from MBS on SUTs. 
	It is also noticeable that integration of JT-NOMA can achieve higher perceived QoE than non-JT-NOMA \cite{QoE_TWC} due to providing opportunities for SUTs to be associated with multiple SBSs. 
	\subsubsection{Audio Service Assessment}
\begin{figure*}\centering
		\begin{tabular}{lccccc}
			\hspace{-0.1cm}\includegraphics[width=6.5cm,height=4cm]{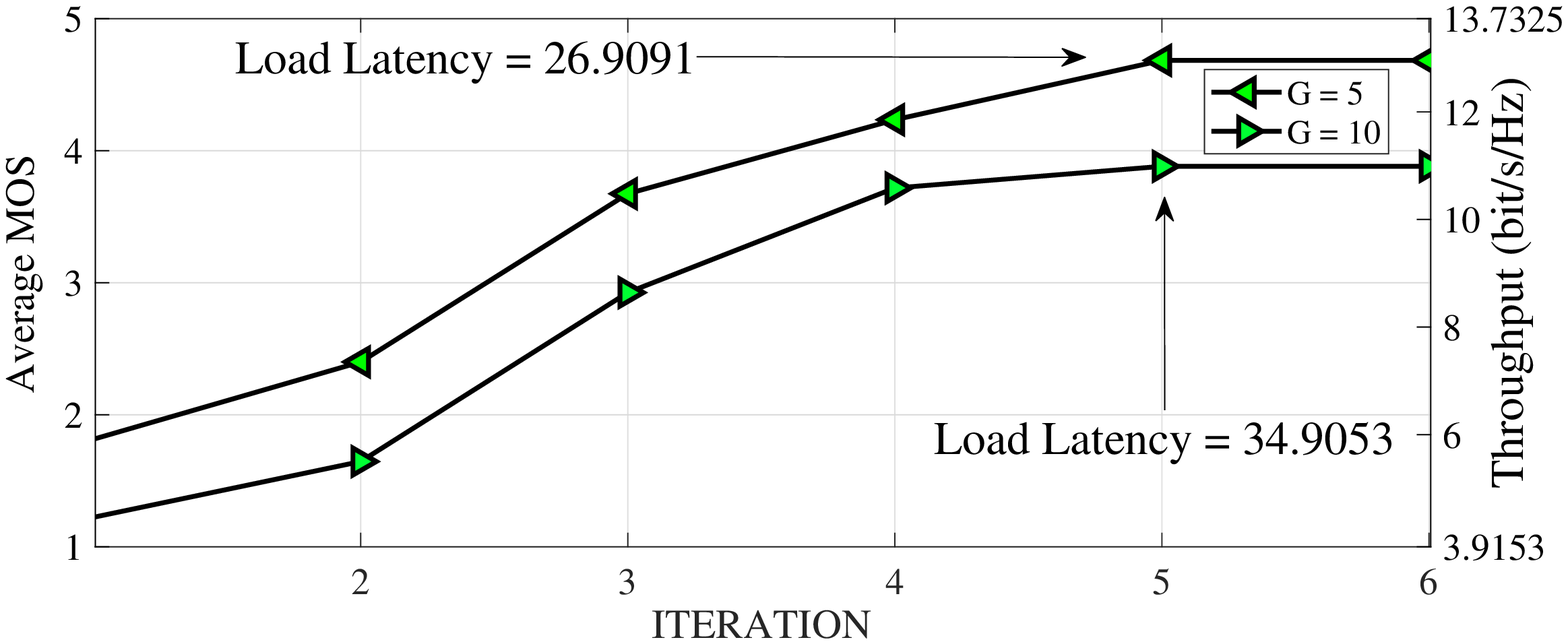}&\hspace{-0.80cm}\includegraphics[width=6.5cm,height=4cm]{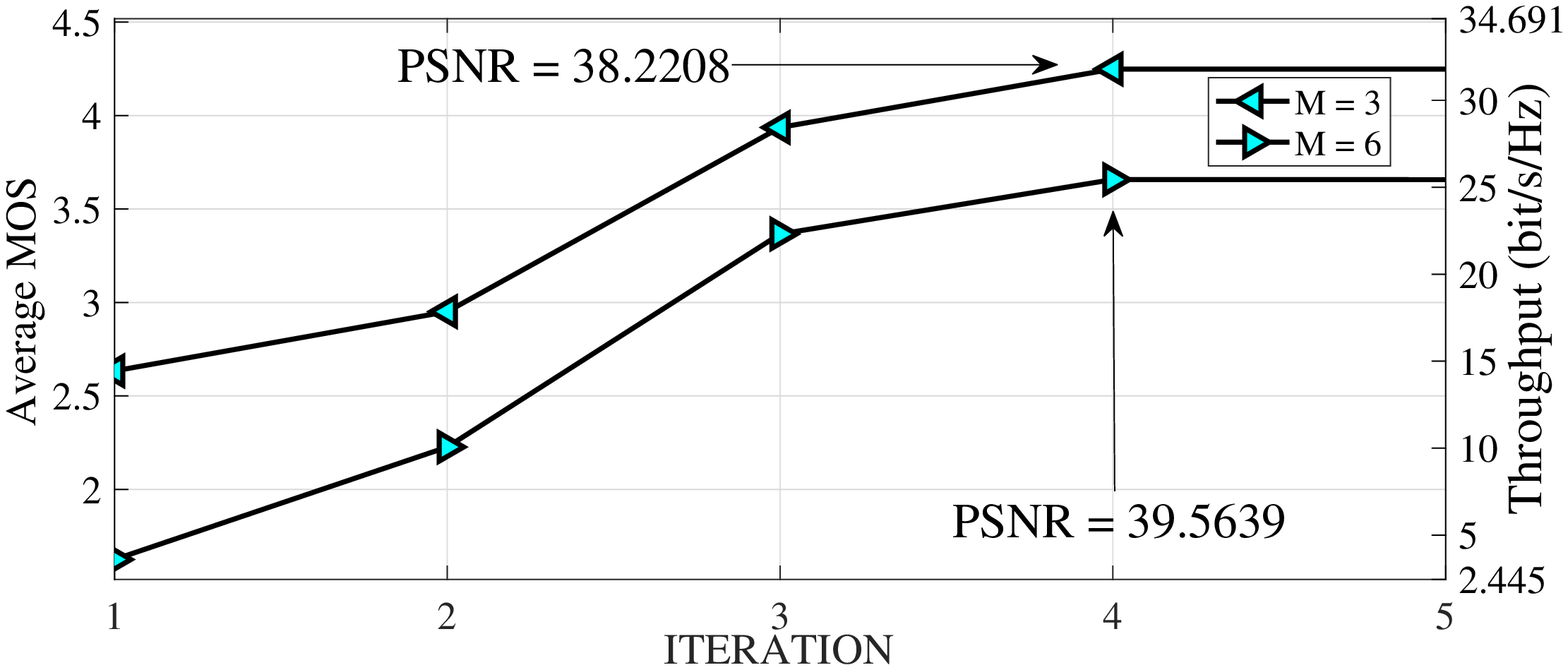}&\hspace{-0.8cm}\includegraphics[width=6.0cm,height=4cm]{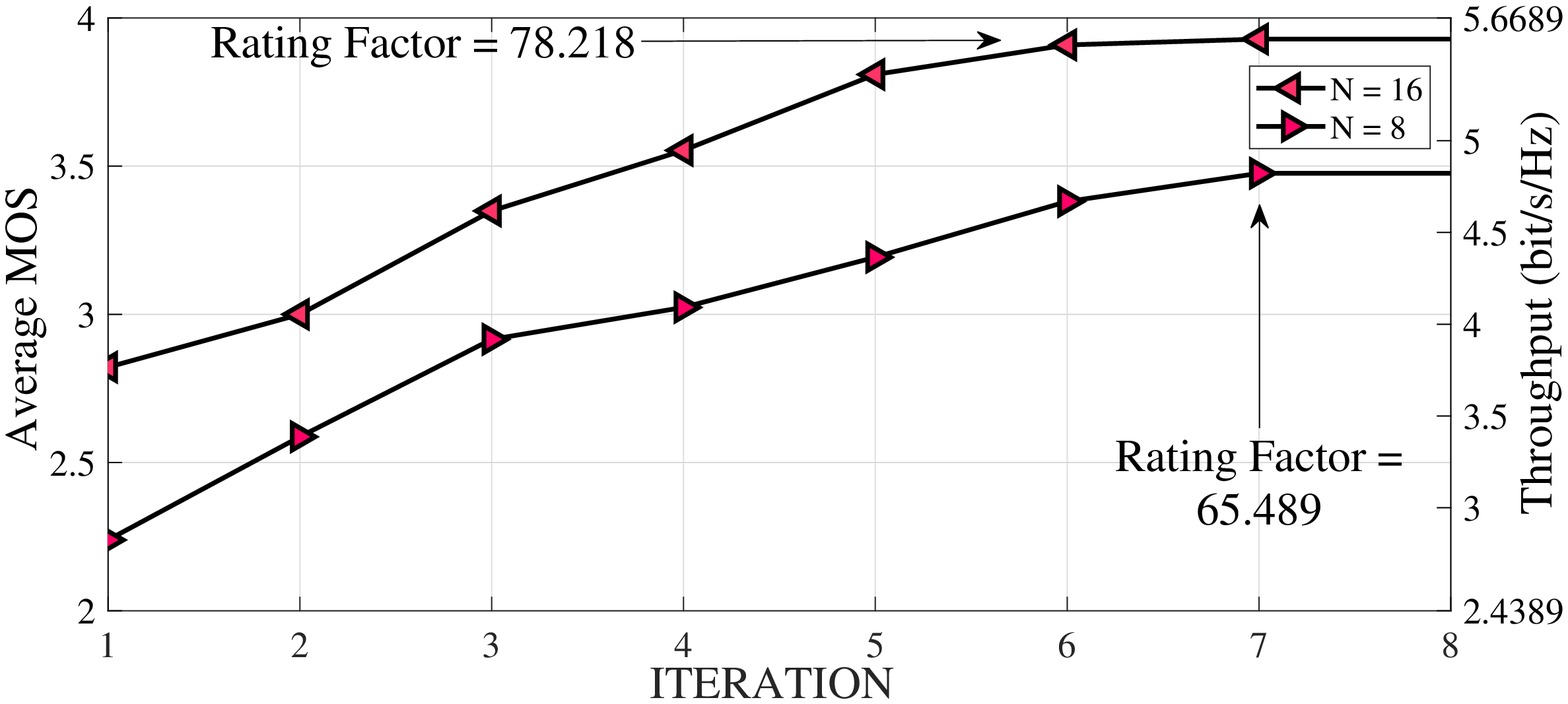}\\
			(a) Web service convergence & (b) Video service convergence & (c) Audio service convergence\\
			\hspace{-0.1cm}\includegraphics[width=6.2cm,height=3.85cm]{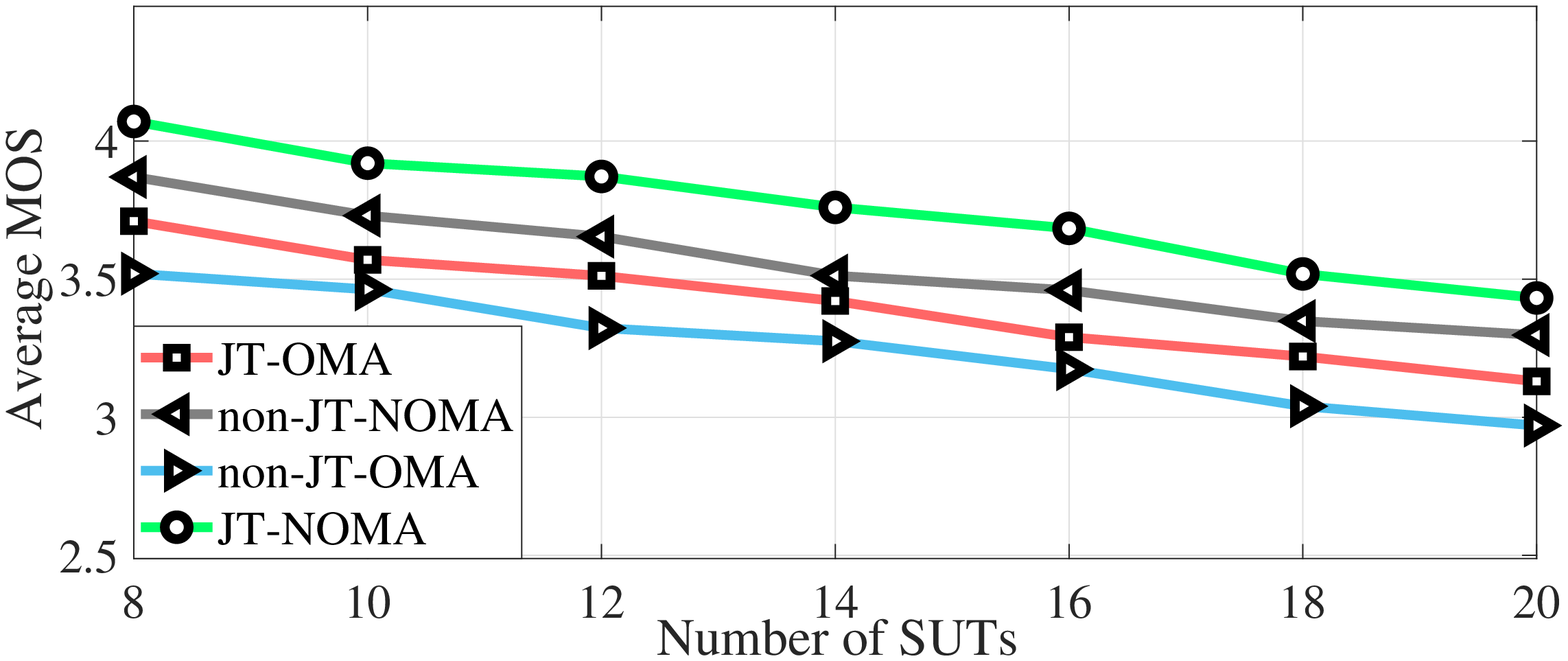}&\hspace{-1.1cm}\includegraphics[width=6.0cm,height=3.80cm]{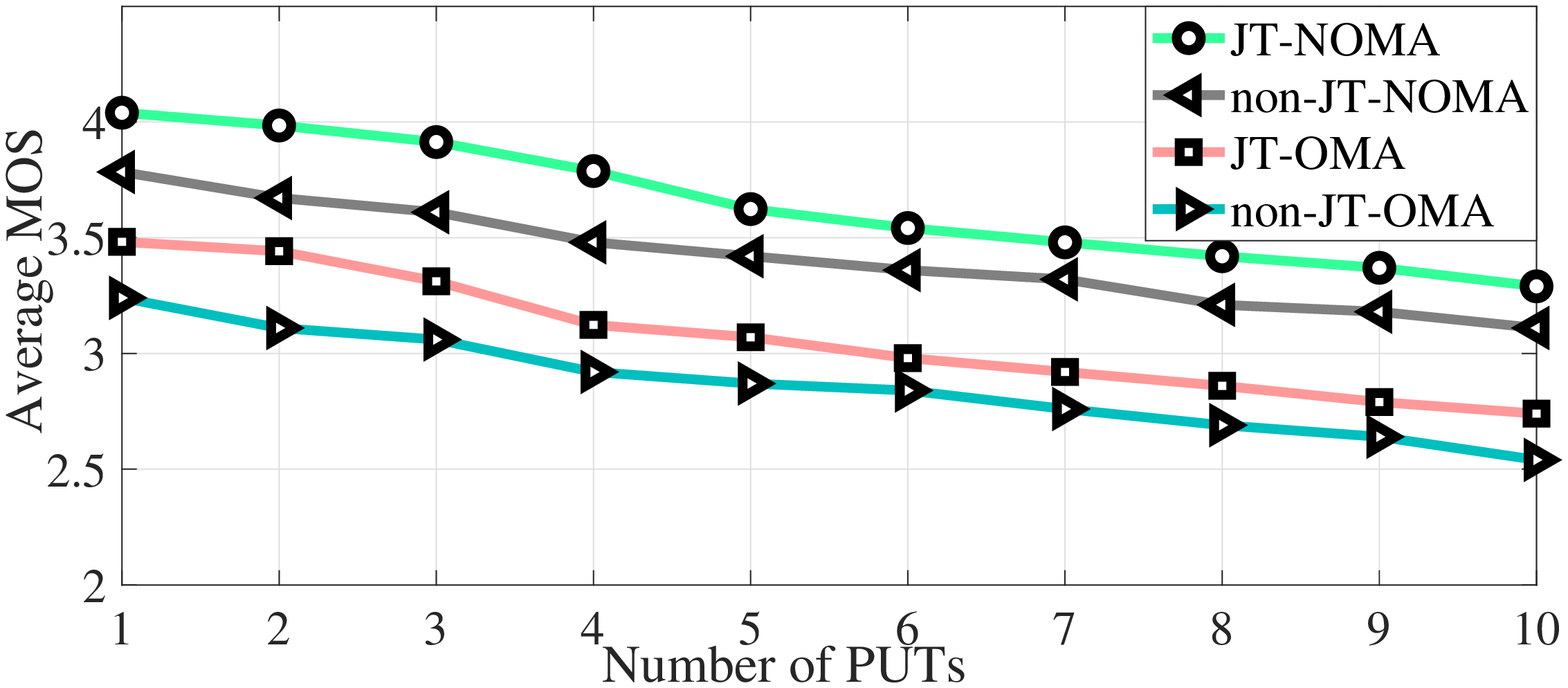}&\hspace{-0.8cm}\includegraphics[width=6.0cm,height=3.8cm]{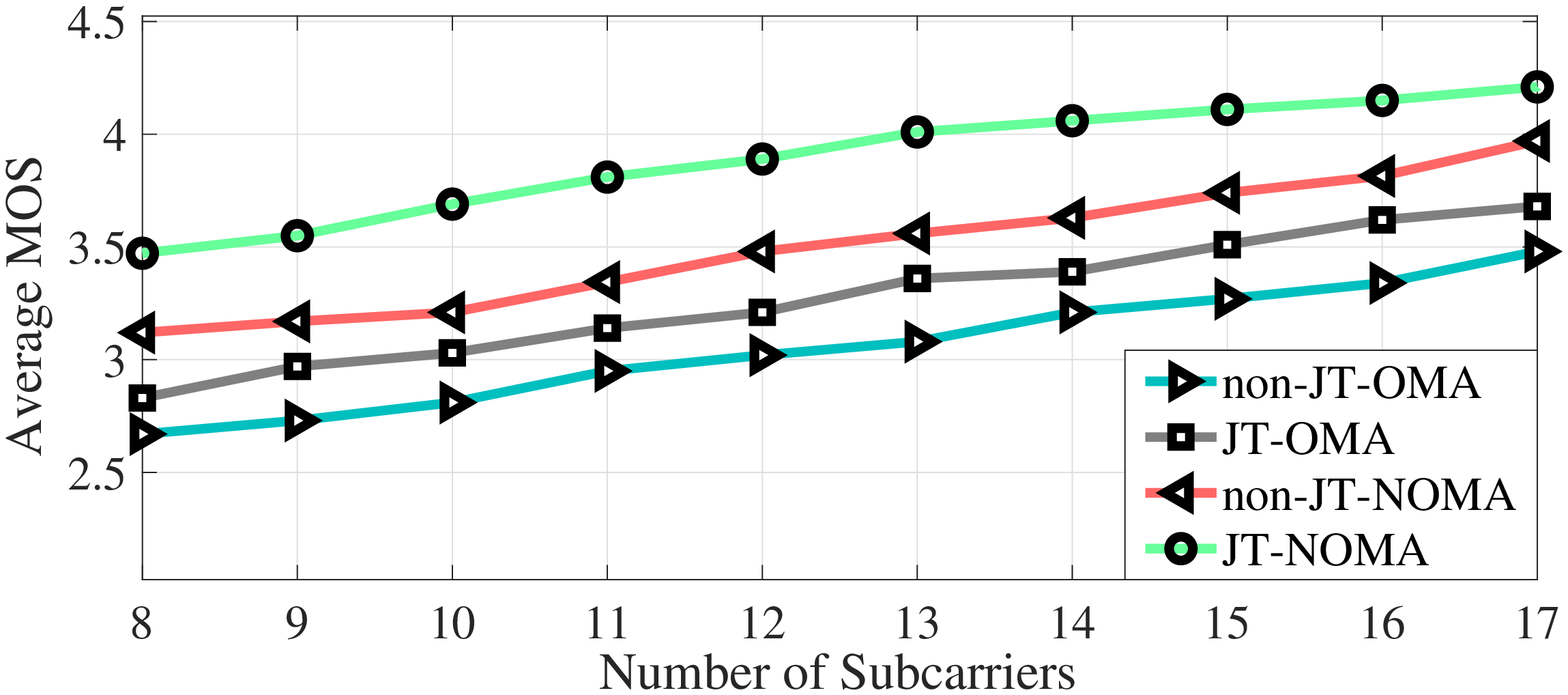}\\
			(d) Web service performance &
			(e) Video service performance & (f) Audio service performance\\
		\end{tabular}
		\caption{Simulation figures for multimedia services} 
		\label{fig:Comp}
	\end{figure*}

	In this section, two cases of maximum and minimum value for Packet Loss Ratio (PLR) is considered as $PLR$ = 20\% and minimum $PLR$ = 0\% respectively\cite{QoE_TCCN}. The network configuration parameters are set to $L$ = 10, $G$ = 8, and $M$ = 4.
	Fig. \ref{fig:Comp}(c) shows the convergence behavior of proposed algorithm for two values of the number of subcarriers with related rating factor\cite{VoIP} in each case is calculated.
	In Fig. \ref{fig:Comp}(f), increasing the number of subcarriers let SUTs access more subcarriers based on carrier aggregation and also more reusing opportunity which leads to higher average MOS per SUT. Subcarriers as the key factor for JT let more scheduling opportunity for SUTs and yield more perceived QoE for JT-enabled baselines in comparison with non-JT. Also, more subcarriers facilitate reuse of the spectrum with respect to NOMA and thus, superiority of the NOMA over OMA is clearly convincing. 
	\section{Conclusions}
	In this paper, we studied the integration of JT and MC-NOMA in the downlink of backhaul limited multi-cell two-tier CRN  to enhance the perceived QoE for web, video, and audio multimedia services. Inspired of various interference categories initiated by reusing the subcarriers by JT-NOMA in CRN context, adequate resource allocation techniques were invoked to the non-liner form of interference terms as well as to overcome the non-convexity of QoE serving functions. In simulation results, the convergence behaviour and superiority of the proposed algorithm is shown using different network parameters for multimedia services.

%




\begin{thebibliography}{99}
\bibitem{QoE-2}
\enquote{Definition of quality of experience,} Int. Telecommun. Union,
Geneva, Switzerland, ITU-T Recommendation TD 109rev2 (PLEN/12),
pp. 16–25, 2007.

\bibitem{NOMA_award}
Z. Ding, Z. Yang, P. Fan and H. V. Poor, \enquote{On the performance of non-orthogonal multiple access in 5G systems with randomly deployed users,} \textit{IEEE Signal Process. Lett.}, vol. 21, no. 12, pp. 1501-1505, Dec. 2014.

\bibitem{NOMA_MC}Y. Sun, D. W. K. Ng, Z. Ding and R. Schober, ``Optimal joint power and subcarrier allocation for MC-NOMA systems," \textit{Proc. IEEE. GLOBECOM,} 2016, pp. 1-6

\bibitem{NOMA_CRN}
I. Budhiraja, S. Tyagi, S. Tanwar, N. Kumar and M. Guizani, \enquote{Cross layer NOMA interference mitigation for femtocell users in 5G environment,} \textit{IEEE Trans. Veh. Technl.}, vol. 68, no. 5, pp. 4721-4733, May. 2019.

\bibitem{JT_NOMA}
M. S. Ali, E. Hossain and D. I. Kim, \enquote{Coordinated multipoint transmission in downlink multi-cell NOMA systems: models and spectral efficiency performance,} \textit{IEEE Wireless Commun.}, vol. 25, no. 2, pp. 24-31, Apr. 2018.

\bibitem{LC}
F. Khoramnejad and M. Rasti and H. Pedram, and E. Hossain,
\enquote{On resource management in load-coupled OFDMAA networks,}
\textit{IEEE Trans. Commun.}, vol. 66, no. 5, pp. 2295--2311, May. 2018

\bibitem{backhaul_1}
Q. Han, B. Yang, G. Miao, C. Chen, X. Wang and X. Guan, \enquote{Backhaul-aware user association and resource allocation for energy-constrained hetNets,} \textit{IEEE Trans. Veh. Technl.}, vol. 66, no. 1, pp. 580-593, Jan. 2017.

\bibitem{JT2}
H. H. M. Tam, H. D. Tuan, D. T. Ngo, T. Q. Duong and H. V. Poor, \enquote{Joint load balancing and interference management for small-cell heterogeneous networks with limited backhaul capacity,} \textit{IEEE Trans. Wireless Commun.}, vol. 16, no. 2, pp. 872-884, Feb. 2017.

\bibitem{QoE_TVT}
H. Abarghouyi, S. M. Razavizadeh and E. Björnson, \enquote{QoE-aware beamforming design for massive MIMO heterogeneous networks,} \textit{IEEE Trans. Veh. Technl.}, vol. 67, no. 9, pp. 8315-8323, Sep. 2018.

\bibitem{QoE_TCCN}
J. Chen, Y. Deng, J. Jia, M. Dohler and A. Nallanathan, \enquote{Cross-layer QoE optimization for D2D communication in CR-enabled heterogeneous cellular networks,} \textit{IEEE Trans. Cogn. Commun. Netw.}, vol. 4, no. 4, pp. 719-734, Dec. 2018.

\bibitem{QoE_TWC}
J. Cui, Y. Liu, Z. Ding, P. Fan and A. Nallanathan, \enquote{QoE-based resource allocation for multi-cell NOMA networks,} \textit{IEEE Trans. Wireless Commun.}, vol. 17, no. 9, pp. 6160-6176, Sep. 2018.

\bibitem{JT7}
F. Baccelli and A. Giovanidis,
\enquote{A stochastic geometry framework for analyzing schedulewise-cooperative cellular networks,} \textit{IEEE Trans. Wireless Commun.}, vol. 14, no. 2, pp. 794--808, 2015.

\bibitem{mmw}
M. Moltafet, R. Joda, N. Mokari, M. R. Sabagh and M. Zorzi, \enquote{Joint access and fronthaul radio resource allocation in PD-NOMA-based 5G networks enabling dual connectivity and CoMP,} \textit{IEEE Trans. Commun.}, vol. 66, no. 12, pp. 6463-6477, Dec. 2018.

\bibitem{CA2}
H. Boostanimehr and V. K. Bhargava, \enquote{Unified and distributed QoS-driven cell association algorithms in heterogeneous networks,} \textit{IEEE Trans. Wireless Commun.}, vol. 14, no. 3, pp. 1650-1662, Mar. 2015.

\bibitem{ALM1}
C. Shen, T. H. Chang, K. Y. Wang, Z. Qiu, and C. Y. Chi, \enquote{Distributed robust multicell coordinated beamforming with imperfect CSI: An ADMM approach,} \textit{IEEE Trans. Signal Process.}, vol. 60, no. 6, pp. 2988-3003, Jun. 2012.

\bibitem{khalili2}
A. Khalili, S. Akhlaghi, H. Tabassum and D. W. K. Ng, \enquote{Joint user association and resource allocation in the uplink of heterogeneous networks,} \textit{IEEE Wireless Commun. Lett.} vol. 9, no. 6, pp. 804–808, Jun. 2020.


\bibitem{Schuber}D. W. K. Ng, E. S. Lo and R. Schober, \enquote{Energy-efficient resource allocation in multi-cell OFDMA systems with limited backhaul capacity,} \textit{IEEE Trans. Wireless Commun.}, vol. 11, no. 10, pp. 3618-3631, Oct. 2012.

\bibitem{optimization}S. Boyd and L. Vandenberghe, \textit{Convex Optimization.} Cambridge University Press, 2004.

\bibitem{Khalili}
A. Khalili, S. Zarandi and M. Rasti, \enquote{Joint resource allocation and offloading decision in mobile edge computing,} \textit{IEEE Commun. Lett.}, vol. 23, no. 4, pp. 684-687, Apr. 2019.
\bibitem{TWC_Ata}A. Khalili, M. Robat Mili, M. Rasti, S. Parsaeefard, and D. W. K. Ng, ``Antenna selection strategy for energy efficiency maximization in uplink OFDMA networks: A multi-objective approach," \textit{IEEE Trans. Wireless Commun.,} vol. 19, no. 1, pp. 595-609, Jan. 2020, 
	\bibitem{Khalili_G1}
	S. Zarandi,  A. Khalili, M. Rasti, and H. Tabassum. "Multi-objective energy efficient resource allocation and user association for in-band full duplex small-cells." \textit{IEEE Trans. on Green Commun. and Networking}, (2020), to appear.
	\bibitem{Khalili_G2}
	A.~Khalili,~S.~Zarandi,~M.~Rasti,~and E.~Hossain,~"Multi-objective optimization for energy- and spectral-efficiency tradeoff in in-band full-duplex (IBFD) communication," \textit{Proc.~IEEE Globecom},~Waikoloa,~HI,~USA,~2019,~pp.~1-6,~Dec.~2019.
\bibitem{VoIP}
M. S. Mushtaq, A. Mellouk, B. Augustin and S. Fowler, \enquote{QoE power-efficient multimedia delivery method for LTE-A,} \textit{IEEE Systems J.}, vol. 10, no. 2, pp. 749-760, Jun. 2016.
\bibitem{WKh}A. Khalili and D. W. K. Ng, "Energy and spectral efficiency tradeoff in OFDMA networks via antenna selection strategy,"\textit{Proc. IEEE WCNC}, Seoul, Korea (South), 2020, pp. 1-6

\end{thebibliography}
\end{document}